# Physics in the Real Universe: Time and Spacetime

George F R Ellis[1]
Mathematics Department, University of Cape Town

**Abstract:** *The Block Universe idea, representing spacetime as a fixed whole, suggests the flow of time is an illusion: the entire universe just is, with no special meaning attached to the present time. This paper points out that this view, in essence represented by usual space-time diagrams, is based on time-reversible microphysical laws, which fail to capture essential features of the time-irreversible macro-physical behaviour and the development of emergent complex systems, including life, which exist in the real universe. When these are taken into account, the unchanging block universe view of spacetime is best replaced by an evolving block universe which extends as time evolves, with the potential of the future continually becoming the certainty of the past; spacetime itself evolves, as do the entities within it. However this time evolution is not related to any preferred surfaces in spacetime; rather it is associated with the evolution of proper time along families of world lines.*

## 1. The Block Universe

The standard spacetime diagrams used in representing the nature of space and time present a view of the entire spacetime, with no special status accorded to the present time; indeed the present (`now') is not usually even denoted in the diagram. Rather all possible `present times' are simultaneously represented in these diagrams on an equal basis. This is the usual space-time view associated both with special relativity (when gravity is negligible, see e.g. Ellis and Williams 2000) and with general relativity (when gravity is taken into account, see e.g. Hawking and Ellis 1973). When the Einstein field equations have as the source of curvature either a vacuum (possibly with a cosmological constant) or simple matter (e.g. a perfect fluid, an electro-magnetic field, or a scalar field), everything that occurs at earlier and later times is locally known from the initial data at an arbitrary time, evolved according to time reversible local physics; hence there is nothing special about any particular time. In a few cases time irreversible physics is taken into account (for example nucleosynthesis in the early universe), but the notion of the present as a special time is still absent.

This view can be formalised in the idea of a *block universe* (Mellor 1998, Savitt 2001, Davies 2002)[2]: space and time are represented as merged into an unchanging spacetime entity, with no particular space sections identified as the present and no evolution of spacetime taking place. The universe just is: a fixed spacetime block. In effect this representation embodies the idea that time is an illusion: it does not `roll on' in this picture. All past and future times are equally present, and there is nothing special about the present (`now'). There are Newtonian, Special Relativity, and General Relativity versions of this view (see Figures 1-4), the latter being most realistic as it is both relativistic and includes gravity.[3]

---

[1] Email address: ellis@maths.uct.ac.za
[2] And see Wikipedia: http://en.wikipedia.org/wiki/Block_time for a nice introduction.
[3] We do not consider here the possible variants when quantum gravity is taken into account.

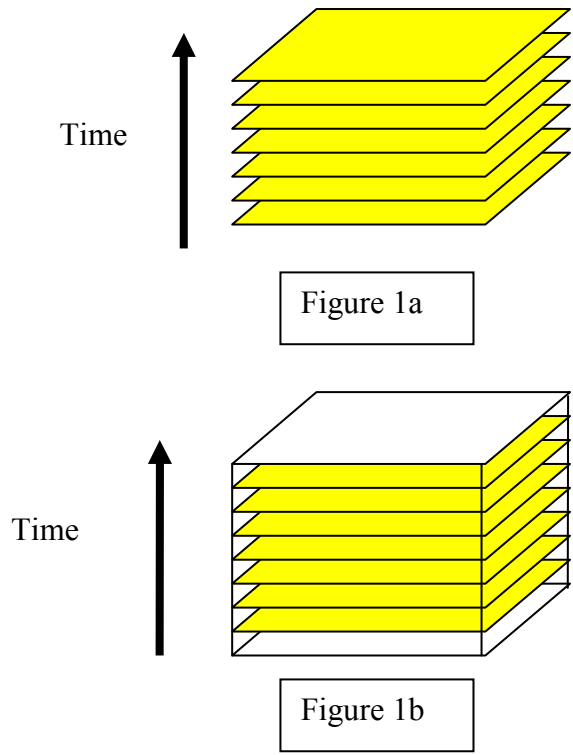

**Figure 1**: *Newtonian spaces (Figure 1a) stacked together to make a Newtonian space-time (Figure 1b) (see Ellis and Williams 2000). Thus this is like plywood: the grain out of which it is constructed remains as preferred space sections (i.e. surfaces of constant time) in spacetime. Data on any of these surfaces determines the evolution of physics to the whole spacetime (through time reversible microphysics).*

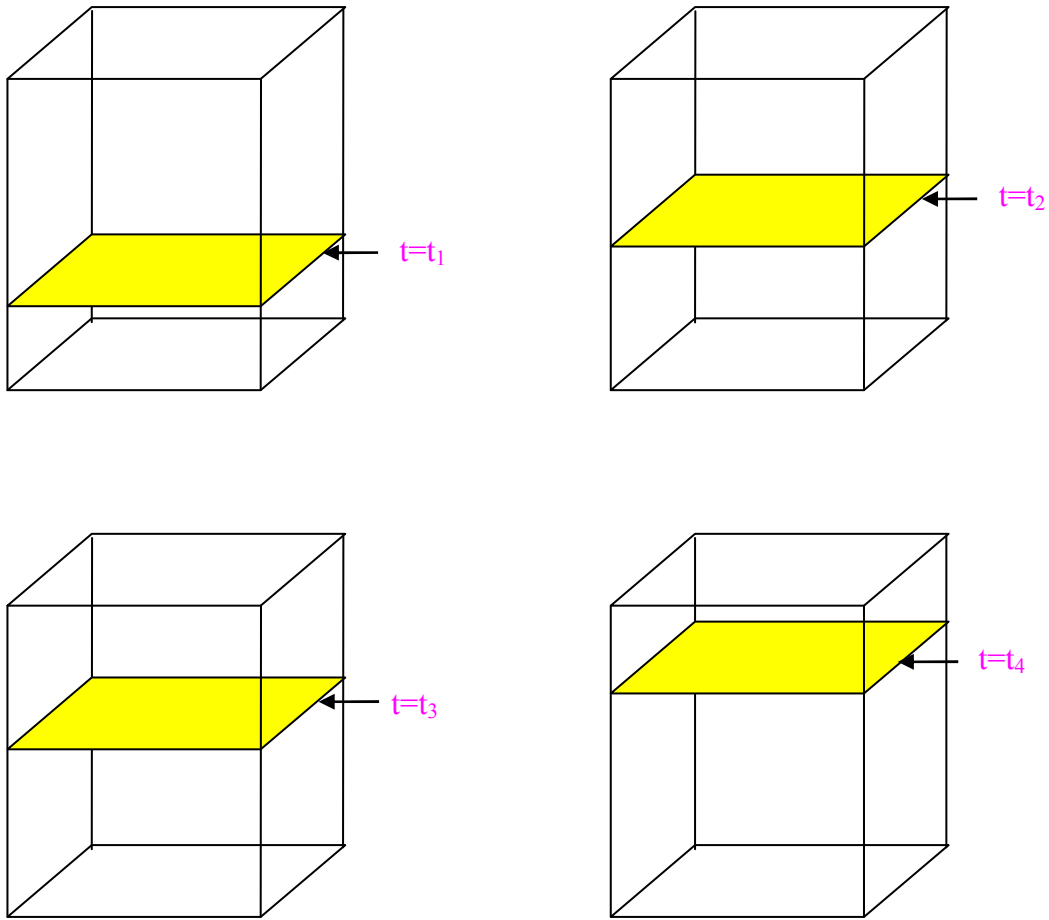

**Figure 2:** *Different parallel time surfaces in a special relativity "block space-time" where spacetime is a block with a set of spatial sections indicating different specific times. These are just different slicings of the same immutable spacetime, but they are not engrained in its structure; it is like a block of glass with no preferred sections. Data on any of these time surfaces determines the evolution of physics to the whole spacetime (through time reversible microphysics).*

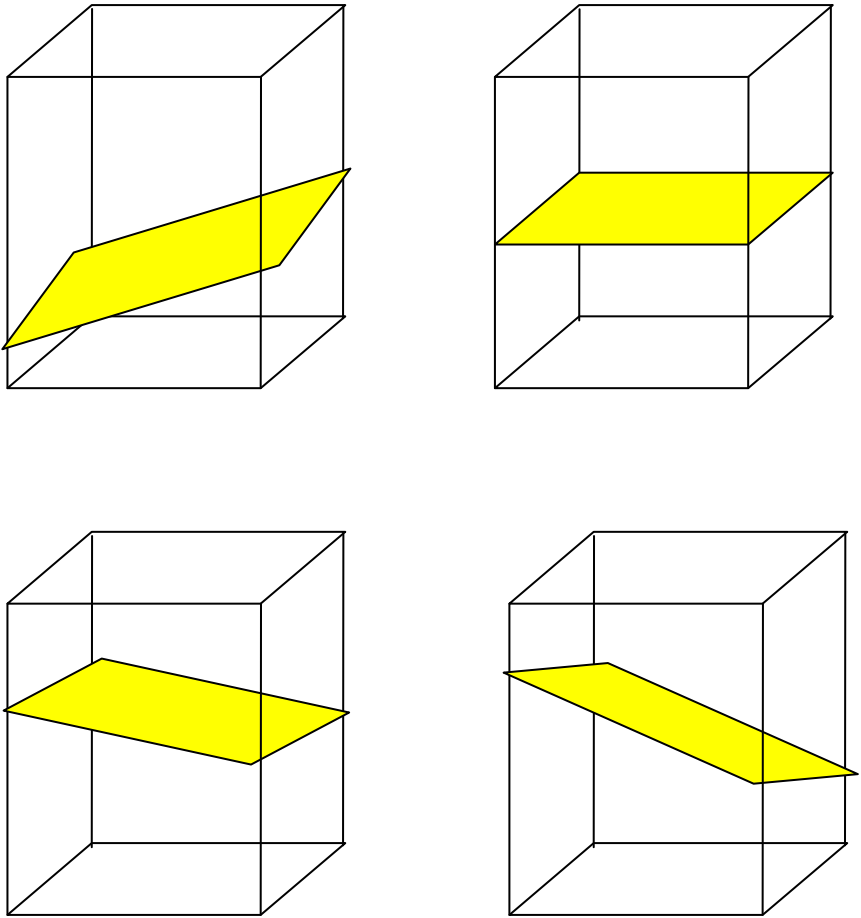

**Figure 3:** *Time surfaces in a special relativity "block space-time" are not unique (Ellis and Williams 2000, Lockwood 2005): they depend on the motion of the observer. Many other families can be chosen, and "space" has no special meaning. Spacetime is the basic object. This is one of the justifications for the block universe picture: we can slice this immutable spacetime in many ways.*

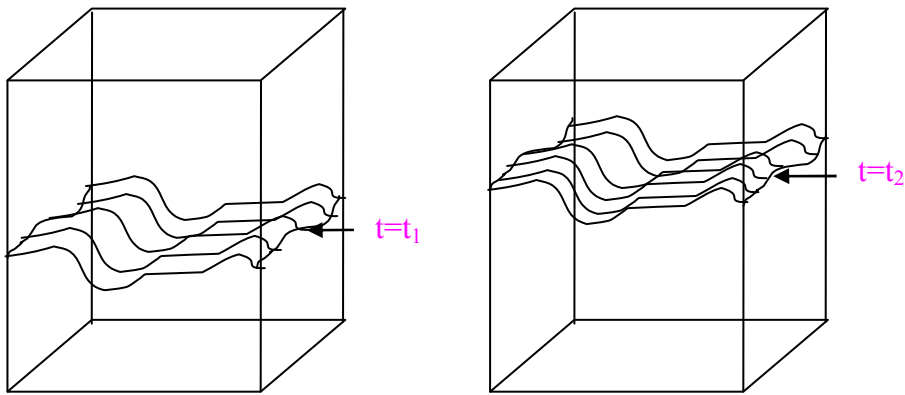

**Figure 4:** *Different time surfaces in a curved block space-time. General relativity allows any `time' surfaces that intersect all world lines locally. The spacetime itself is also curved. Future and past physics, including the spacetime itself, are locally determined from the data on any such surface.*

The warrant for this view in the case of special relativity is the existence and uniqueness theorems for the relevant fields on a fixed Minkowski background spacetime; for example, the existence and uniqueness theorems for fluid flows, for Maxwell's equations, or for the Klein Gordon equation (see Hadamard 1923; Wald 1984: 243-252). In the case where gravity is significant, the warrant is the existence and uniqueness theorems of general relativity for suitable matter fields (Hawking and Ellis 1973: 226-255; Wald 1984: 252-267). They show that for such matter, initial data at an arbitrary time determines all physical evolution, including that of the space-time structure, to the past and the future equally, because we can predict and retrodict from that data up to the Cauchy horizon. The present time has no particular significance; it is just a convenient time surface we chose on which to consider the initial data for the universe. We could have equally chosen any other such surface.

**2. The Unfolding of Time**

This block view is however an unrealistic picture because it does not take complex physics or biology seriously; and they do indeed exist in the real universe. The irreversible flow of time is one of the dominant features of biology, as well as of the physics of complex interactions and indeed our own human experience (Le Poidevin 2004). Its associated effects are very significant on small scales (cells to ecosystems), though they are probably unimportant on large scales (galaxies and above).

This scale-dependence is a key feature, intimately related to the question of *averaging scales* in physics: every description used in any physical theory, including every spacetime description, involves an explicit or implicit averaging scale (Ellis 1984, Ellis and Buchert 2005). To deal adequately with complex structures in a spacetime, one must be very clear what averaging scale is being used. The flow of time is very apparent at some scales (that of biology for example), and not apparent at others (e.g. that of classical fundamental physics).

Classical micro-physics is time-reversible: detailed predictability to the past and future is in principle possible. It is in this case that `the present' may be claimed to have no particular meaning. However the numbers of interactions involved, together with the existence of chaotic systems, can make detailed prediction impracticable in practice, leading to the use of statistical descriptions: we can predict the kinds of things that will happen, but not the specific outcome. Time-irreversible macro-physics and biology is based in the micro-physics, but with emergent properties that often involve an overt `flow of time' and associated increase of entropy. The past is fixed forever, and can in principle be largely known; the future is unknown and mostly unpredictable in detail. The present is more real than the undetermined future, in that it is where action is now taking place: it is where the uncertain future becomes the immutable past. Various views are possible on how to relate these different aspects:

> "There have been three major theories of time's flow. The first, and most popular among physicists, is that the flow is an illusion, the product of a faulty metaphor. The second is that it is not an illusion but rather is subjective, being deeply ingrained due to the nature of our minds. The third is that it is objective, a feature of the mind-independent reality that is to be found in, say, today's scientific laws, or, if it has been missed there, then in future scientific laws … Some dynamic theorists argue that the boundary separating the future from the past is the moment at which that which was undetermined becomes determined, and so "becoming" has the same meaning as "becoming determined." (Fieser and Dowden 2006, Section 7.)

The first and second views are those associated with the block universe picture and usual space-time diagrams. Can one find a spacetime view supporting the third position? Yes one can; this is what we present below. Before doing so we first look more closely at a realistic view of the physics involved.

**2.1 A Broken Wine Glass, Coarse-graining**
A classic example of an irreversible process is the breaking of a wine glass when it falls from a table to the floor (Penrose 1989: 304-309). Now the key point for my argument here is that the precise outcome (the specific set of glass shards that result and their positions on the ground) is unpredictable: as you watch it fall, you can't foretell what will be the fragmentation of the glass. You can't predict what will happen at this level of detail because a macro-description of the situation (the initial shape, size, position, and motion of the glass) does not have enough detail of its micro-properties (defects in its structure, for example) to work this out.

The underlying physics is deterministic but our classical predictive model of what happens is not. It just might be possible to determine the fracturing that will occur if you have a detailed description of the crystalline structure of the glass, but that data – which in any case would be extremely difficult to obtain - is not available in a macro-description. From the macro viewpoint what happens is random; you can give a statistical prediction of the likely outcome, but not a detailed definite prediction of the unique actual outcome. From a classical micro viewpoint it is deterministic - you just need enough data and computing power to find out what will happen; but the macro-description and associated spacetime picture does not contain that detailed information. *You can only find out what happens by watching it happen*; the physical result (the specific shapes and positions of the shards, for example) unfolds as time progresses. Furthermore, this lack of predictability holds in both time directions. Considering the backward direction of time, you can't reconstruct the details of the process of destruction (what happened when) from the fragments on the ground, because you can't tell when the glass fell by looking at the resulting fragments. Even if we accessed all the micro-data available at late times, that uncertainty would remain: no amount of data collection will resolve it, once the thermal traces of the fall have dissipated and merged into the background noise.

Similar results will hold for example for the explosion of a bomb: the distribution of fragments of a bomb that will occur is not predictable from macro-data available by external observation (see Figure 5).

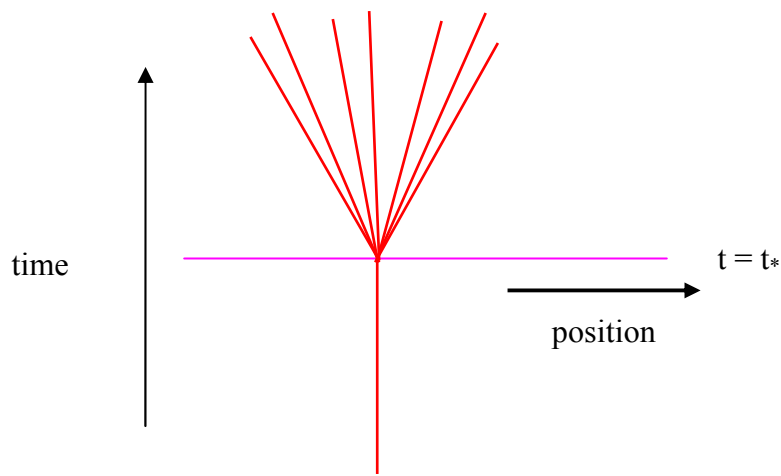

**Figure 5**: *The impossibility of prediction in the real world: space time diagrams of the explosion of a bomb at time $t = t_*$. The macro state description for $t < t_*$ does not determine the number of fragments and their motion for $t > t_*$.*

Generally in relating the description of a physical system at different scales, the microscopic data $(x_i, p_j)$ needed for the detailed phase space ($x_i$ are position coordinates and $p_i$ the corresponding momenta) is coarse-grained to give macroscopic variables $(X_i, P_j)$ characterising the phase space associated with the averaged macro-description of the system. If the averaging operator is A, then

$$A: (x_i, p_j) \longrightarrow (X_i, P_j). \qquad (1)$$

Many different micro states correspond to the same macrostate, which is the source of entropy (Penrose 1989: 309-314). The micro-dynamics, usually given by Hamilton's equations (Penrose 1989:175-184), is governed by an operator $\varphi_t$ giving the change over the time interval *t*:

$$\varphi_t: (x_i, p_j) \longrightarrow (x_i', p_j') = (\varphi_t x_i, \varphi_t p_j). \qquad (2)$$

and the macro dynamics by a corresponding operator $\Phi_t$. The micro variables will be subject to dynamic and structural constraints:

$$C_1(x_i, p_j) = 0 \quad (\partial C_i/\partial p_j \neq 0), \quad C_2(x_i) = 0, \qquad (3)$$

the latter describing the structural relations of the system (for example, the crystal structure of the glass). These constraints are preserved by the dynamics (2):

$$C_1(x_i', p_j') = 0, \quad C_2(x_i') = 0. \qquad (4)$$

However in some cases, different initial micro states that correspond to the same initial macro state result in different final macro outcomes, because the dynamics and averaging do not commute:

$$A \varphi_t \neq \Phi_t A. \qquad (5)$$

Then detailed micro level predictability determines the macro level outcomes but does not lead to reliable emergent macro level behaviour: indeed then $\Phi_t$ is not well defined (see Ellis 2006b: Figure 5). This will be the case for example when chaotic behaviour occurs (Thomson and Stewart 1987). Furthermore the structural constraints only hold for a limited range of values of the dynamic variables: if these bounds are exceeded, the constraints will be violated:

$$C_2(x_i') \neq 0 \qquad (6)$$

(a glass breaks or bomb goes off as in the examples above, or a phase change takes place). The way this happens is not described by the dynamics (2), which assume these constraints are preserved, and is invalid otherwise. A dynamical analysis is required that covers the change of constraints and resulting new dynamics; but even phase changes from water to ice are not fully predictable at present (Laughlin 2005: Chapter 4).

**2.2 Friction, Coarse-graining**

In general, friction effects mean we have an inability to retrodict if we lose information below some level of coarse graining. The simplest example is a block of mass $m$ sliding on a plane, slowing down due to constant limiting friction $F = -\mu R$ where $\mu$ is the coefficient of friction and $R = mg$ is the normal reaction, where g is the acceleration due to gravity (Spiegel 1967). The motion is a uniform deceleration; if we consider the block's motion from an initial time $t = 0$, it comes to rest at some later time $t_* > 0$. For $t < t_*$ the velocity v and position x of the object are given by

$$v_1(t) = -\mu g\, t + v_0, \quad x_1(t) = -\tfrac{1}{2}\mu g\, t^2 + v_0 t + x_0 \qquad (7)$$

where $(v_0, x_0)$ are the initial data for $(v,x)$ at the time $t = 0$. This expression shows that it comes to rest at $t_* = \mu g / v_0$. For $t > t_*$, the quantities v and x are given by

$$v_2(t) = 0, \quad x_2(t) = X \text{ (constant)}, \qquad (8)$$

where $X = -\frac{1}{2} \mu g\, t_*^2 + v_0 t_* + x_0$.

The key point now is that from the later data (8) at any time $t > t_*$ you cannot determine the initial data $(v_0, x_0)$, nor the time $t_*$ when the object came to rest, thus you cannot reconstruct the trajectory (7) from that data. You cannot even tell if the block came from the left or the right (see Figure 6).

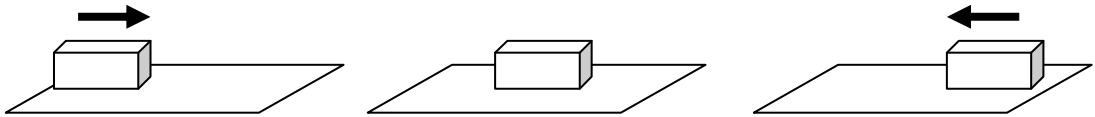

**Figure 6a:** *The impossibility of retrodiction in the real world: a block sliding on a surface. The stationary block in the centre might have come from the left, and stopped under friction, or from the right. Observing it at rest does not tell us which was the case.*

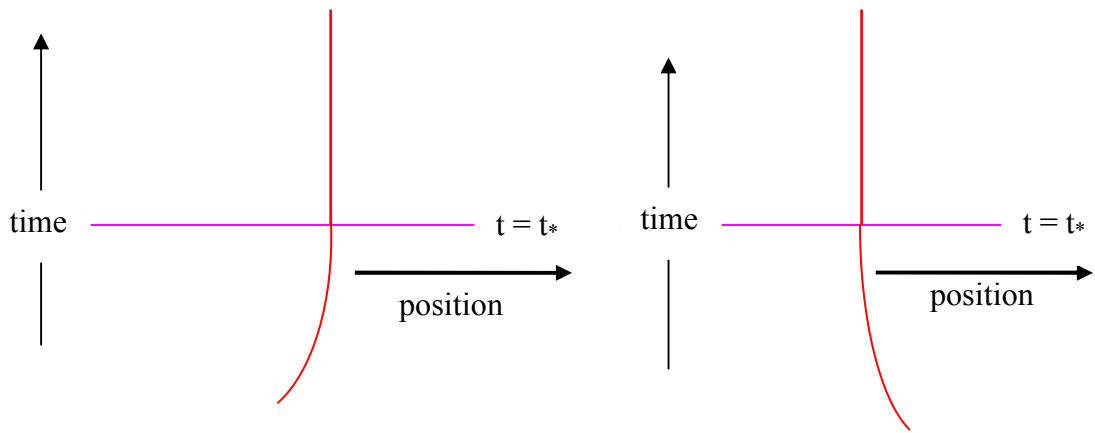

**Figure 6b**: *Space time diagrams of the position of the block for the two cases illustrated above. The state for $t > t_*$ does not determine the state for $t < t_*$, or even determine the value $t_*$*

If you could measure the distribution of heat in the table top soon enough after the block stopped, through thermal imaging for example, you could work out what had happened (because it's motion will have been converted into heat). Thus the inability to retrodict soon after the block comes to rest is again a result of using a macro picture that does not include all the detailed data: here, the thermal motion of particles in the table top (which will then dissipate away and be lost in the environment; this data too will soon become irretrievably lost). This is of course the essence of successful physics models: using a simplified picture that throws most of the detailed data away, and concentrating on essentials. The cost is that the ability of the model to predict is strictly limited.

Similar results hold for any viscous processes or dissipative system: the final state will generically be an attractor; once it has settled down you cannot tell from macro data what initial state the system came from - it could have been any point in the basin of attraction (Thompson and Stewart, 1987).

**2.3 Quantum Uncertainty**
In these examples, our inability to predict is associated with a lack of detailed information. So if we fine-grained to the smallest possible scales and collected all the available data, could we then determine uniquely what is going to happen? No, we can't predict to the future in this way because of foundational quantum uncertainty relations (see e.g. Feynman 1985, Penrose 1989, Isham 1997), apparent for example in radioactive decay (we can't predict precisely when a nucleus will decay and what the velocities of the resultant particles will be) and the motion of a stream of particles through a pair of slits onto a screen (we can't predict precisely where a photon or electron will end up on the screen). It is a fundamental aspect of quantum theory that this uncertainty is unresolvable: *it is not even in principle possible to obtain enough data to determine a unique outcome of quantum events.* This unpredictability is not a result of a lack of information: it is the very nature of the underlying physics.

More formally: if a measurement of an observable $A$ takes place at time $t = t_*$, initially the wave function $\psi(x)$ is a linear combination of eigenfunctions $u_n(x)$ of the operator $\tilde{A}$ that represents $A$: for $t < t_*$, the wave function is

$$\psi_1(x) = \Sigma_n \psi_n u_n(x). \tag{9}$$

(see e.g. Isham 1997: 5-7). But immediately after the measurement has taken place, the wave function is an eigenfunction of $\tilde{A}$: it is

$$\psi_2(x) = a_N u_N(x) \tag{10}$$

for some specific value N. The data for $t < t_*$ do not determine the index N; they just determine a probability for the choice N. One can think of this as due to the probabilistic time-irreversible collapse of the wave function (Penrose 1989: 260-263). Invoking a many-worlds description (see e.g. Isham 1997) will not help: in the actually experienced universe in which we make the measurement, N is unpredictable. Thus the initial state (9) does not uniquely determine the final state (10); and this is not due to lack of data, it is due to the foundational nature of quantum interactions. You can predict the statistics of what is likely to happen but not the unique actual physical outcome, which unfolds in an unpredictable way as time progresses; you can only find out what this outcome is after it has happened. Furthermore, in general *the time $t_*$ is also not predictable from the initial data*: you don't know when `collapse of the wave function' (the transition from (9) to (10)) will happen (you can't predict when a specific excited atom will emit a photon, or a radioactive particle will decay).

*We also can't retrodict to the past at the quantum level*, because once the wave function has collapsed to an eigenstate we can't tell from its final state what it was before the measurement. You cannot retrodict uniquely from the state (10) immediately after the measurement takes place, or from any later state that it then evolves to via the Schrodinger equation at later times $t > t_*$, because knowledge of these later states does not suffice to determine the initial state (9) at times $t < t_*$: the set of quantities $\psi_n$ are not determined by the single number $a_N$.

The fact that such events happen at the quantum level does not prevent them from having macro-level effects. Many systems can act to amplify them to macro levels, including photomultipliers[4] (whose output can be used in computers or electronic control systems). Quantum fluctuations can change the genetic inheritance of animals (Percival 1991) and so influence the course of evolutionary history on Earth. Indeed that is in effect what occurred when cosmic rays[5] – whose emission processes are subject to quantum uncertainty - caused genetic damage in the distant past:

> "The near universality of specialized mechanisms for DNA repair, including repair of specifically radiation induced damage, from prokaryotes to humans, suggests that the earth has always been subject to damage/repair events above the rate of intrinsic replication errors ….. radiation may have been the dominant generator of genetic diversity in the terrestrial past" (Scalo et al 2001).[6]

Consequently *the specific evolutionary outcomes on life on Earth (the existence of dinosaurs, giraffes, humans) cannot even in principle be uniquely determined by causal evolution from conditions in the early universe, or from detailed data at the start of life on Earth.* Quantum uncertainty prevents this, because it significantly affected the occurrence of radiation-induced mutations in this evolutionary history. The specific outcome that actually occurred was determined as it happened, when quantum emission of the relevant photons took place: the prior uncertainty in their trajectories was resolved by the historical occurrence of the emission event, resulting in a specific photon emission time and trajectory that was not determined beforehand, with consequent damage to a specific gene in a particular cell at a particular time and place that cannot be predicted even in principle.

**2.4 Space-time curvature: Time-dependent Equations of State**
So far we have considered unpredictability of the evolution of local systems in a fixed space time; this needs to be taken into account in our space-time pictures of such interactions. But can this uncertainty affect the nature of space time itself? Yes indeed; in general relativity theory, matter curves space time, and the curvature of spacetime then affects the motion of matter (Hawking and Ellis 1973, Misner et al 1973, Wald 1984). We can have unpredictability at both stages of the non-linear interaction that determines the future spacetime curvature.

First, as regards matter determining spacetime curvature: can unpredictable local processes have gravitational effects that in turn affect space-time curvature? Hermann Bondi (1965) posited a pair of orbiting massive objects (`Tweedledum' and `Tweedledee') where internal batteries drive motors slowly altering the shape of the bodies from oblate to prolate spheroids and back, thereby changing their external gravitational field. This enables exchange of energy and information between them via gravitational induction: time-dependent terms in the field equations and Bianchi identities are negligible. Time variation in the source term in the constraint equations conveys energy from one body to the other by altering the electric part $E_{ab}$ of the Weyl conformal curvature tensor (the `free gravitational field') in the intervening spacetime[7]. This then changes the tidal source term $E_{ab}$ in the deviation equation[8] for matter in the second body, altering its shape. Information can be conveyed between them by altering the time pattern of these variations: computer control of the motors allows an arbitrary signal to be transferred between Tweedledum and Tweedledee, that cannot be predicted from the initial data for the gravitational field (it is specified by the computer programme). This unpredictability is a result of the implicit coarse-grained description of the physical system: changes in space time curvature occur that cannot be predicted from external view of the objects because that description does not include details of the internal mechanisms, including the specific bits making up the stored computer programme (these would be represented at a much finer level of description). One can have similar processes involving gravitational radiation. Consider two identical spherical masses at the end of a strong rod, able to turn about a vertical axis. An electric motor rotates the rod, and is controlled via a computer to turn the rod at high angular speeds $\omega_i$ for a series of time intervals $T_i$ separated by stationary intervals $t_i$. This creates a time-dependent gravitational dipole that will emit gravitational waves according to standard formulae (Misner et al 1973: Ch.36), with oscillations in the electric and magnetic parts of the Weyl tensor carrying energy and information from one place to another during each interval $T_i$. This time-dependent field can in principle be detected by a distant gravitational wave detector.

---
[4] See Wikipedia: http://en.wikipedia.org/wiki/Photomultiplier for their physical realisation.
[5] See http://www.chicos.caltech.edu/cosmic_rays.html for a brief summary of their origin.
[6] See also for example (Babcock and Collins 1929, Rothschild 1999, National Academy 2005).

Second, the motion of matter can be affected in a similar manner. Suppose we attached a large number of massive rocket engines to one side of the Moon and fired them simultaneously. This would change the orbit of the Moon (for a while its motion would be non-geodesic) in a way that is cumulative with time. This then would affect the way it curved spacetime in the future, for its future position relative to the Earth would be different from what it would otherwise have been. The local Weyl tensor will have been altered and so tides on the Earth would be altered. Thus such engineering efforts can change the future space-time curvature and its physical effects. Again computer control allows an arbitrary time evolution to be specified.

Generically the point is that *explicitly time dependent equations of state can affect the future development of spacetime, and how this will work out is unpredictable from the macro initial data at any specific time*. If the spacetime description were detailed enough to include the classical mechanisms involved in such physical causation (the clocks, computers, motors, rocket engines, etc. that caused such changes, including the computer programme) then they might be predictable; but a standard spacetime picture does not include this detailed data. Furthermore, *such mechanisms could include a random element making such prediction impossible even in principle*. One might for example arrange for the computer programme to use as input, signals from either a particle detector that detects the particles emitted through radioactive decay of unstable atoms, or a photon detector that responds to individual photons from a distant quasar. Then quantum uncertainty in the particle emission process would prevent precise prediction of the future spacetime curvature, even in principle.[9] In effect we would in these cases be amplifying quantum uncertainty to astronomical scales.

---

[7] See Ellis (1971) for the relevant equations (the `div E' Bianchi identities).
[8] The generalisation of the geodesic deviation equation to non-geodesic motion.
[9] It has been suggested to me that in view of these outcomes, one should query the uncertainty principle of quantum theory, seeking a deterministic version instead, as suggested for example by David Bohm and David Gross. Such a theory may indeed become widely accepted one day, and then what is

Human intentionality underlies the unpredictable functioning of the mechanisms (motors, computers, etc.) considered above, as they would be the result of human agency (implied by their supposed existence as designed objects). However this kind of effect can occur in other contexts without human intervention, indeed it has already happened in the expanding universe at very early times. According to the standard inflationary model of the very early universe, we cannot predict the specific large-scale structure existing in the universe today from data at the start of the inflationary expansion epoch, because density inhomogeneities at later times have grown out of random quantum fluctuations in the effective scalar field that is dominant at very early times:

> "Inflation offers an explanation for the clumpiness of matter in the universe: quantum fluctuations in the mysterious substance that powered the [inflationary] expansion would have been inflated to astrophysical scales and therefore served as the seeds of stars and galaxies" (Hinshaw 2006).[10]

Thus *the existence of our specific Galaxy, let alone the planet Earth, was not uniquely determined by initial data in the very early universe*. The quantum fluctuations that are amplified to galactic scale by this process are unpredictable in principle.

## 2.5 Overall: a lack of predictability in the real universe

In summary, *the future* is not determined till it happens because of[11]
   (a) time-dependent equations of state, which can be information driven,
   (b) quantum uncertainty, which can be amplified to macro scales,
and also in practice by
   (c) statistics / experimental errors / classical fluctuations,
amplified by
   (d) chaotic dynamics /occurrence of catastrophes.
Essentially all realistic models of the universe except for very large scale cosmology are non-deterministic. Sufficient reasons for this are:

(i) coarse-graining by its very nature introduces a statistical element; and

(ii) quantum processes occur on the small scale, and can be amplified to macro scales, so there is no deterministic microscopic model from which fully predictive classical macroscopic models can always be derived.

Because of these effects, we cannot predict uniquely to the future from present-day data; indeed its detailed features remain open and causally undetermined by initial conditions. Statistical prediction is however possible in contexts where emergent complexity, and in particular biological agency, is unimportant (it is not necessarily useful when Darwinian selection effects or human agency are active, for example predicting the probability of existence of giraffes or ostriches at present, or of specific endangered species in the future). This determines the kinds of thing that will happen, but not the specific outcome that actually occurs.

---

presented here would have to be revised; until this happens, the prudent view is to provisionally accept one of the major outcomes of physics last century, namely that quantum uncertainty is indeed real.
[10] See Kolb and Turner 1990 or Dodelson 2003 for details.
[11] Another possibility is effects due to emergent complexity, including animal and human agency. Although this is an important topic (see for example Ellis 2005a, 2005b, 2006a), it will not be pursued further in this paper.

*The past* has happened and is fixed, so the nature of its existence is quite different than that of the indeterminate future. However we cannot causally retrodict uniquely to the past from present day initial data by using the appropriate evolution equations for the matter, because of friction and other dissipative effects (see Section 2.2) and the quantum measurement process (`collapse of the wave function', see Section 2.3). What we can do, is observe present-day features resulting from past events (geological and archaeological data, photographic images, written records of past events, etc), and thereby attempt to determine what in fact occurred by analysing these observations in conjunction with the dynamic projection of local physics from present data to the past (Lockwood 2005: 233-256).[12]

## 3. A Realistic Space-Time Picture

The time reversible picture of fundamental physics underlying the block universe viewpoint simply does not take these kinds of phenomena into account, specifically because it does not take cognisance of how complex phenomena arise from the underlying micro-physics, with the emergence of the arrow of time. It does not take seriously the physics and biology of the real world but rather represents an idealised view of things which is reasonably accurate on certain (very large) scales where very simplified descriptions are successful.

In order to take the physical situations considered in the previous section into account, we need to modify the block universe pictures so as to adequately represent causation in these contexts. How do we envisage spacetime and the objects in it as time unrolls? A way to do this is to consider an *Evolving Block Universe* (`EBU') model of reality, with spacetime ever growing and incorporating more events as time evolves along each world line.

To motivate this, consider the following scenario (Figure 7): A massive object has rocket engines attached at each end to make it move either left or right. The engines are controlled by a computer that decides what firing intervals are utilised alternately by each engine, on the basis of a non-linear time dependent transformation of signals received from a detector measuring particle arrivals due to random decays of a radioactive element. These signals at each instant determine what actually happens from the set of all possible outcomes, thus determining the actual spacetime path of the object from the set of all possible paths (Figure 8). This outcome is not determined by initial data at any previous time, because of quantum uncertainty in the radioactive decays.[13] As the objects are massive and hence cause spacetime curvature, the spacetime structure itself is undetermined until the object's motion is determined in this way. Instant by instant, the spacetime structure changes from indeterminate (i.e. not yet determined out of all the possible options) to definite (i.e. determined by the specific physical processes outlined above). Thus a definite spacetime structure comes into being as time evolves. It is unknown and unpredictable before it is determined.

---

[12] There is uncertainty as regards both the future and the past, but its nature is quite different in these two cases. The future is uncertain because it is not yet determined: it does not yet exist in a physical sense. Thus this uncertainty has an ontological character. The past is fixed and unchanging because it has already happened, and the time when it happened cannot be revisited; but our knowledge about it is incomplete, and can change with time. Thus this uncertainty is epistemological in nature.

[13] In effect this diagram shows the multiple options of the Everett-Wheeler branching universe view (Isham 1997), but with specific choices made as the wave function collapses time after time (Penrose 1989), resulting in an emerging unique outcome as time progresses.

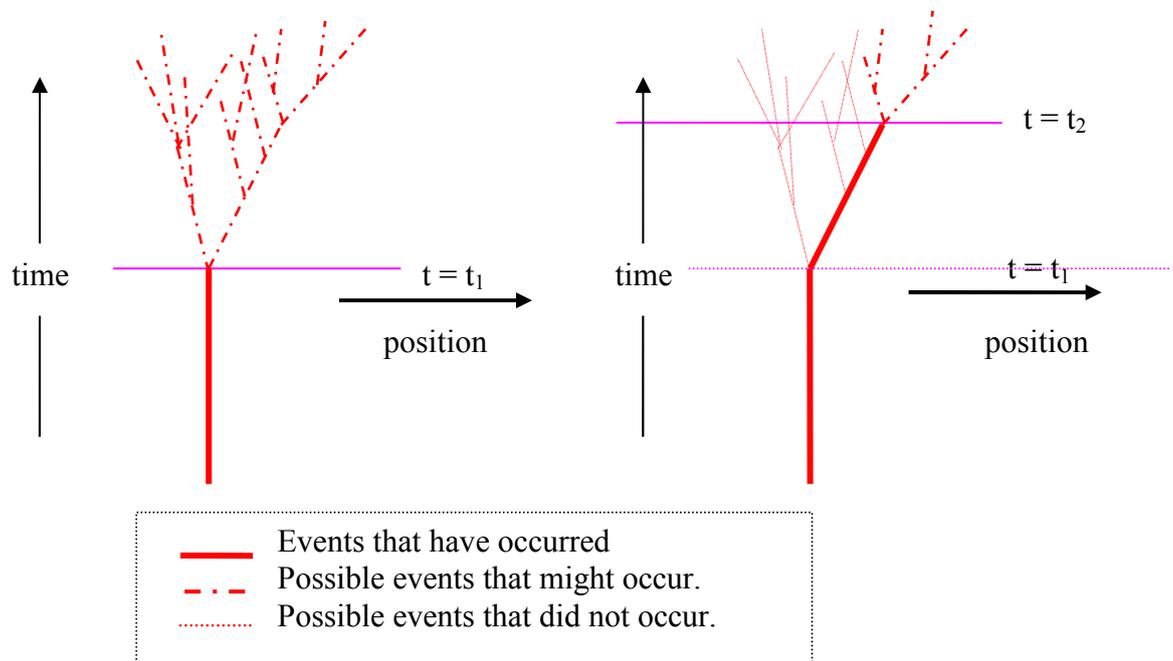

**Figure 7**: *Motion of a particle world line controlled in a random way, so that what happens is determined as it happens. On the left events are determined till time* $t_1$ *but not thereafter; on the right, events are determined till time* $t_2 > t_1$, *but not thereafter.*[14]

The Evolving Block Universe model of spacetime represents this kind of situation, showing how time progresses, events happen, and history is shaped. Things could have been different, but second by second, one specific evolutionary history out of all the possibilities is chosen, takes place, and gets cast in stone. This idea was proposed many years ago (Broad 1923)[15], but has not caught on in the physics community.[16]

We now consider it successively in the contexts of Newtonian theory, special relativity, and general relativity.

**3.1 The Newtonian Case**
Here we consider events in spacetime as evolving from indefinite to determinate as time passes; the past is fixed and immutable, and hence has a completely different status than the future, which is still undetermined and open to influence. The kinds of `existence' they represent are quite different: the future only exists as a potentiality rather than an actuality. The existential nature of the present is indeed unlike that of the past or future, for it is the set events we can (according to Newtonian theory) actually influence at any instant in our history.

---

[14] See Lockwood (2005), Figure 1.1 (page 12).
[15] For quotes from Broad's book in this regard, see Ted Sider's notes at
http://fas-philosophy.rutgers.edu/sider/teaching/415/HO_growing_block.pdf .
[16] Tooley (1997) also puts forward such a theory in an interesting way, but (unlike what is presented here) proposes modifying special relativity for this purpose. This is clearly unlikely to gain acceptance.

We can represent this through a growing spacetime diagram with unique time surfaces (Figure 8), the passing of time marking how things change from being indefinite (and so not yet existing) to definite (and so having come into being), with the present marking the instant when we can act and change things.

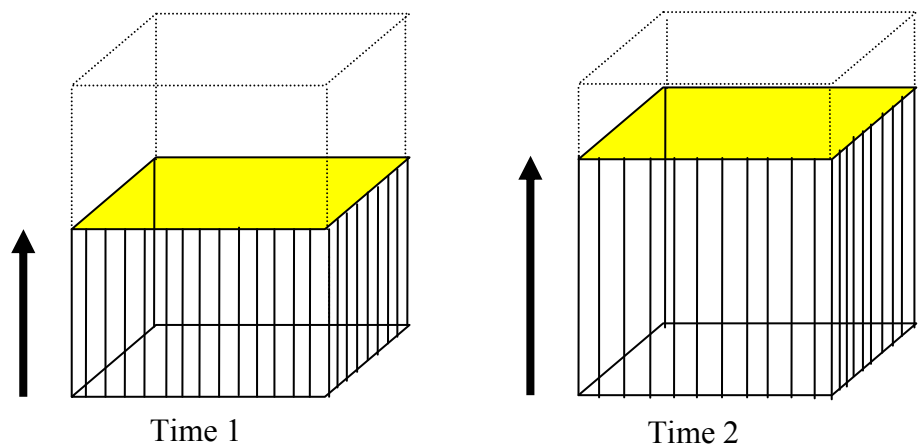

Time 1                    Time 2

**Figure 8:** *Time represented as evolving in a universe where events emerge as choices are made. The future is indeterminate, the past is fixed; thus they have a completely different status. The determinate part of spacetime extends into the future as time evolves and potentiality becomes reality (through time irreversible macrophysics with underlying quantum uncertainty). However the nature of the future spacetime is already fixed before events in it evolve, in the cases of Newtonian theory and special relativity.*

In this case we can associate the passing of time uniquely with the preferred spatial sections of Newtonian space-times, representing Newton's absolute time. `The present' exists and is unique. However *although the events in it are uncertain, the nature of the future to-be spacetime is known and immutable*. We do not have to engage in uncertain prediction in order to know what it will be.

**3.2 Special Relativity**
This is like the Newtonian case: we represent the past of each event as fixed and immutable, but the future nature of events as still undetermined. The present is where uncertainty about events changes to certainty. And as in Newtonian theory, the *nature* of the future spacetime is known and immutable (it is just Minkowski spacetime), even though the events in it are unknown.

However the time surfaces are no longer invariant under change of reference frame: they depend on the observer's motion relative to the coordinate system (see Figure 3). So the usual objection to the idea of a special relativistic evolving universe is, *How do we choose which surfaces are associated with the evolution of spacetime*? This choice is arbitrary, and so the unfolding of time is indeterminate: it is not a well defined unique physical process. We need to turn to a world-line based picture, which is natural in general relativity, to get an answer.

**3.3 General Relativity**
In this case, the present is again represented as where the indeterminate nature of potential physical events changes to a definite outcome, but now *even the nature of the future spacetime is taken to be uncertain until it is determined at that time,* along with the physical events that occur in it. A further major feature is that because spacetime is curved, unlike the special relativity case, *in particular solutions of the Einstein equations there are in general geometrically and physically preferred spacelike surfaces and timelike world lines*, related to the specific physics of the situation. These represent broken symmetries in the solutions to the Einstein field equations: the solutions have less symmetry than the equations of the theory

One can suggest that in this case, the transition from present to past does not take place on specific spacelike surfaces; rather *it takes place pointwise at each spacetime event*. The implicit "now" of figures 8 and 9 (or of any "flowing time" concept) is replaced by a "here-now" (space-time point), and for both the "now" and the "here-now" the past is determined (exists relative to the [here]-now), the future is undetermined, and the [here]-now is a moment of passage from one state to the other.

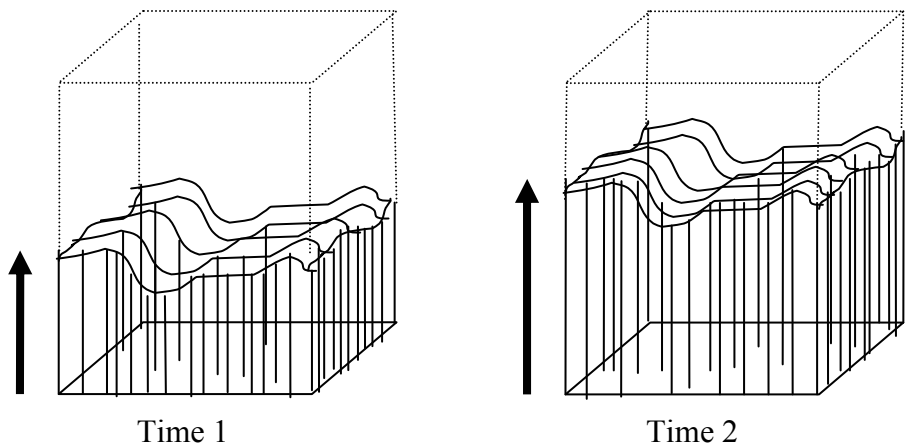

Time 1                    Time 2

**Figure 9:** *An evolving curved space-time picture that takes macro-phenomena seriously. Time evolves along each world line, extending the determinate spacetime as it does so (what might be changes into what has happened; indeterminate becomes determinate). The particular surfaces have no fundamental meaning and are just there for convenience (we need coordinates to describe what is happening). You cannot locally predict uniquely to either the future or the past from data on any `time' surface (even though the past is already determined). This is true both for physics, and (consequently) for the spacetime itself: the developing nature of spacetime is determined by the evolution (to the future) of the matter in it.*

However the constraints on what future can emerge at a given here-now are not point-wise constraints but (in relation to any local coordinates) constraints involving spatial derivatives, or, roughly speaking, neighbouring points. So if evolution takes place pointwise, it still involves a degree of spatial coordination between neighbouring points, even though the neighbouring point might not "yet exist" relative to a different here-now until it lies in the past. It is convenient to introduce local coordinates in order to determine how this works, involving a splitting of spacetime into space and time as in the ADM formalism (Arnowitt, Deser, and Misner 1962, Misner et al 1973: 520-528, Anninos 2001), and evolution along the coordinate lines introduced as determined by the shift function. But then it is physically sensible to focus on this feature: that is, while it may in a sense take place pointwise, it is more convenient to consider the *evolution as taking place along timelike world lines*,[17] rather than being determined by any universal time defined by spacelike surfaces. Indeed this is strongly suggested both by the way that time is determined in General Relativity as a path integral along timelike world lines, and also by the nature of the examples discussed in the previous section, where physical effects that determine what happens are focused on changes that take place along histories of matter, represented by timelike worldlines.[18]

But then the question is, which world line should be chosen? The potential problem is the arbitrariness in the choice of the world lines. There seem to be two choices: either
  (**ET1**) *we regard the evolution of time as being allowed to take place along any world lines whatever, none being preferred*, or
  (**ET2**) *the evolution of time takes place along preferred world lines, associated with a symmetry breaking that leads to the emergence of time.*[19]

In the latter case, the question is, which world lines are the key ones that play this crucial physical role? In specific realistic physical situations, there will be preferred world lines associated with the average motion of matter present, as in the case of cosmology (Ellis 1971), and there will be preferred time surfaces associated with them if the matter flow is irrotational. This occurs in particular in the case of the idealised Robertson-Walker models of standard cosmology, where as long as matter is present, there exist uniquely preferred irrotational and shear-free worldlines that are eigenvectors of the Ricci tensor. These then form a plausible best basis for description of physical events and the evolution of matter: there is a unique physical evolution determined along each such a family of world lines with its associated unique time surfaces, which are invariant under the space time symmetries. Then one might propose that the evolution of time is associated with these preferred timelike world lines and perhaps associated spacelike surfaces, being an emergent property associated with the broken symmetries represented by these geometrical features in curved spacetimes.

But there may be several competing such choices of world lines in more realistic cases, for example realistic perturbed cosmological models such as are needed for structure formation studies will have multiple matter components present with differing 4-velocities (see e.g. Dunsby et al 1992). For this reason, we might rather consider the evolution as taking place along *arbitrary families of world lines*, corresponding to the freedom of choice of the shift vector in the ADM formalism for

---

[17] In most real world contexts, even though it is in principle possible to exert causal influences along null lines (i.e. at the speed of light), this is in practice unimportant except for small-scale situations where lasers are important.

[18] In those cases where radiation dominates, as in the early universe before decoupling, the average motion of the radiation is represented by timelike world lines.

[19] And presumably to the arrow of time (Ellis and Sciama 1972, Davies 1974, Zeh 1992).

General Relativity (Arnowitt et al 1962, Misner et al 1973: 520-528, Anninos 2001). A key result then is that no unique choice for these world lines needs to be made in the standard GR situation with simple equations of state; the ADM theory says *we locally get same result for the evolving spacetime, whatever world lines are chosen.* You can choose any time lines you like to show how things will have evolved at different places (that is, on different observer's world lines) at different times (that is, at various proper times along those world lines). But this view has no foundationally preferred status: you could have chosen different world lines, corresponding to different shift vectors, and a different relation between times on the world lines, corresponding to different choices of the laps function; the resulting four dimensional spacetime is the same. In any specific situation, some of those descriptions will be more natural and easier to use and understand than others; but this is just a convenience, and any other surfaces and world lines could have been chosen.

Thus in the classical GR case, we get a consistent picture: things are as we experience them. Time rolls on along each world line; the past events on a world line are fixed and the future events on each world line are unknown. Spacetime grows as in the Newtonian picture, but now even the spacetime structure itself is to be determined as the evolution takes place (Figure 10). The metric tensor determines the rate of change of time with respect to the coordinates, for this is the fundamental meaning of the metric (Hawking and Ellis, 1973). A gauge condition determines how the coordinates are extended to the future. Conservation equations plus equations of state and associated evolution equations determine how matter and fields change to the future, including the behaviour of ideal clocks, which measure the passage of time. The field equations determine how the metric evolves with time, and hence determine the future space-time curvature. The whole fits together in a consistent way, determining the evolution of both space-time and the matter and fields in it,[20] as is demonstrated for simple equations of state by the existence and uniqueness theorems of general relativity theory (Hawking and Ellis 1973: 226-255; Wald 1984: 252-267).

Thus no unique choice needs to be made for the conventional ADM formalism, *which is deterministic;* for these standard theorems assume classical deterministic physics rules the micro-world. But the whole point of this paper is that most models are *not* deterministic, irreversible unpredictable processes and emergent properties will take part in determining space time curvature; on relatively small scales, even human activity does so (when we move massive objects around). A realistic extension of the above will take into account quantum uncertainty in the evolution of the matter and fields, giving a probability for the future evolution of particles, fields, and hence for spacetime, rather than a definite prediction.[21] Quantum evolution will determine the actual outcome that occurs in a probabilistic way (see the examples in Section 2.4). But then the problem is, if two choices of world lines are made *in two different indeterministic futures*, then it is probable that the two evolutions will not agree. In such a scenario we would have something more like Wheeler's "many fingered time" – the different proper times along arbitrary world lines do not knit together to form a global concept of time that is meaningful in determining a unique evolution along all world lines. How then can an evolving block universe emerge from this situation?

---

[20] Smart's objections to an objective passage or flow of time (Smart 1967: 126) are comprehensively answered by Lockwood (2005: 13-18).
[21] This is what is done in the semi-classical calculations of inflationary universe perturbations, for details see Kolb and Turner (1990), Dodelson (2003).

In the following section, we consider how this might work out in the various indeterministic cases discussed above (in Section 2).

**4 The emergence of a block universe**

The fundamental argument in this section is based on a simple observational fact:

(**OF**) *In the real universe domain we actually inhabit, a unique classical space-time structure does indeed emerge at macro scales from the underlying physics.*
This is true for each of the cases discussed in Section 2: in each case, we are indeed able to describe what happens via an evolving block universe. It is for this reason that we are able to regard special and general relativity as successful theories in their appropriate contexts within the local universe domain in which we live. Consequently we will take this as a fundamental observational fact about the real universe.[22]

The implication of (**OF**) is immediate:
(**EB**) *Whatever conditions are needed to imply the existence of an emergent block universe at macro scales, whether related to a particular set of world lines as in* (**ET1**), *or the emergence of the same spacetime whatever world lines are chosen as in* (**ET2**), *are satisfied in our real physical observable universe domain*.
There could be other universe domains or hypothetical universes in which this is not true, in which for example a classical spacetime structure never emerges; but that is not the case on this Earth or indeed, as far as we can tell, within the visible universe.

So even in those cases where we are at present unable to determine why this is the case or indeed what the required conditions are, it seems clear that *whatever integrability or consistency conditions are needed to guarantee the emergence of a growing block universe are in fact satisfied in the real universe we see around us*. That is the basic feature we assume in what follows.

**4.1 Classical cases**
A main cause of indeterminism discussed above is coarse-graining (Sections 2.1 and 2.2 above). Here there is an underlying deterministic theory where one can apply things like existence and uniqueness theorems to evolution of the underlying fields, but this is lost as a result of coarse graining at a macroscopic level.

This picture might lead to a description of an EBU that would be consistent with a (local) pointwise evolution as suggested in section 3. For example, a stochastic macro-evolution could be produced for a statistical ensemble (e.g. canonical) of initial microstates corresponding to a given coarse-grained state, which could lead to a stochastic ADM formulation.[23] However one may also claim that any averaging scheme in fact involves choices of special world lines around which the averaging is defined. Thus this is plausibly consistent with the vision **ET2** of emergence of time (seen at the averaged scale) as taking place along preferred world lines that are intimately bound up with the averaging process.

---

[22] The observable part of the expanding universe, since the time of decoupling of matter and radiation.
[23] I am grateful to a referee for this suggestion.

A second possible cause of indeterminism is a "time-dependent equation of state" (Section 2.3 above), perhaps associated with emergent complexity or human agency. If the equation is time-dependent, the question arises as to which time is the relevant one? In some situations it will be some local time defined along a particular set of worldliness, which will presumably provide the answer to the question as to the worldlines along which the evolution actually takes place. In other situations the time dependence might be given by some foliation, in which case this again would seem to give rise to a natural EBU in terms of the process **ET2** considered above.

**4.2 Quantum indeterminism**

A different form of indeterminism is provided by quantum uncertainty (Sections 2.3 and 2.4 above). The way that this fits in with an EBU will depend very much on the description one gives of quantum processes and measurements. Even at the level of quantum fields on a curved background there will be many technical difficulties in relating this to an EBU and these would be even worse if one attempted to say something about quantum gravity. The situation is also problematic because there is no agreed conceptual formulation of quantum theory in a cosmological context. In this context the "measurements" of section 2.4 cannot be clearly identified, and if one instead use a decoherent histories approach there remain problems of determining the families of supports and related operator algebras that yield decoherent histories. Even the attempt to pose the alternatives (**ET1**) and (**ET2**) may not make sense. On the other hand semiclassical approaches that sidestep these problems are in widespread use, particularly the extension of the standard theory that is used in inflationary theory. This takes quantum uncertainty into account, and is based on a description via preferred time lines and space-like surfaces, and so is compatible with **ET2**.

However from the viewpoint of this section, one does not need to solve these difficult technicalities. Rather one can take a pragmatic approach, as indicated in the examples discussed in sections 2.3 and 2.4, where well-known and proven properties of quantum theory provide the basis of the conclusions, without a need to discuss how they arise from the underlying quantum field theory. There are two alternatives: (1) that (for reasons as yet unknown) a unique classical spacetime necessarily emerges from the underlying quantum theory, because of the nature of that theory and the relevant boundary conditions; (2) that there is no guarantee in general that a good classical spacetime emerges, hence cases occur where spacetime pictures simply cannot be drawn for the real universe that eventuates; in this case, further integrability conditions must be imposed on quantum theory if one is to guarantee this emergence.

In view of the observation (**OF**) above and its consequence (**EB**), it does not matter for present purposes which is the case: *either quantum theory on a curved background does indeed imply unique outcomes for physics and spacetime under the conditions that occur in the universe domain in which we live*, and an evolving block universe is a natural outcome of the physics, or *there are extra integrability conditions that should be imposed on those theories to guarantee that this outcome occurs; and it is then an observational fact these conditions are satisfied in the real universe in which we live, as an Evolving Block Universe does indeed occur.*

Investigating what these conditions are, and how they are satisfied, is a separate issue – essentially the contentious question of the nature of emergence of both classical physics and a classical curved space-time. This will not be tackled here.

**4.3 Global Issues**
Where difficult dynamics issues may well arise is in terms of the global extension of the local results. There may indeed be no globally unique evolution, because this may genuinely depend on choice of families of world lines – different global extensions of a local region may result from different such choices, see for example the case of the Taub-NUT Universe (Hawking and Ellis 1973: 170-78). There may be a need to make a definite choice, motivated by physical considerations, in order to attain a unique global extension. This is an issue that will need investigation.

**4.4 The Far Future Universe**
What is the final fate of this growing universe? It is nothing other than the usual fixed unchanging block universe picture, but understood in a new way as the ultimate state of the evolving block universe (EBU) proposed here. It is *what will be in the far future, when all possible evolution has taken place along all world lines*. It is thus, when properly considered, the *Final Block Universe* (FBU). In this picture of the Far Future Universe, time is no longer an indicator of change that will take place in the future, because it has all already happened. At every event along every world line, the present has been and gone. The implications of the existence of this unchanging final state are very different from those usually imputed to the block universe. Things are unchanging and eternal here not because they are immutably implied by the past, but because they have already occurred. This view represents the final history of the universe when all choices have been made and all alternative histories chosen: when it has been determined that the Earth would come into existence, that specific continents would develop, and that particular types of dinosaurs would emerge and then die out.

**4.5 Summary: An Evolving Block Universe**
Thus the view proposed here is that spacetime is extending to the future as events develop along each world line in a way determined by the complex of causal interactions; these shape the future, including the very structure of spacetime itself, in a locally determined (pointwise) way. It is an Evolving Block Universe that continues evolving along every world line until it reaches its final state as an unchanging Final Block Universe. One might say that then time has changed into eternity. The future is uncertain and indeterminate until local determinations of what occurs have taken place at the space-time event `here and now', designating the present on a world line at a specific instant; thereafter this event is in the past, having become fixed and immutable, with a new event on the world line designating the present. There is no unique way to say how this happens relatively for different observers; analysis of the evolution is conveniently based on preferred (matter related) world lines rather than time surfaces. However in order to describe it overall, it will be convenient to choose specific time surfaces for the analysis, but these are a choice of convenience rather than necessity.

This paper does not attempt an analysis of how this relates to the philosophy of time (see e.g. Markosian 2002, Fieser and Dowden 2006). Rather I just make one remark in this regard at this point:[24] in terms of the metaphysics of time, this view is that of *possibilism* (the tree model), described in Savitt (2001: Section 2.1) and Hunter (2006: Section 3).

---

[24] And see also Section 4.3 below.

# 5 Overall: A more realistic view

The standard block universe picture is based on reversible micro-physics, not realistic irreversible macro-physics. However when coarse-graining and emergent effects such as biology are taken into account, with internal variables leading in effect to highly non-linear time dependent equations of state, time does roll on, indeed this is one of the most fundamental features of our lives: intention changes the future; the past is fixed forever and cannot be changed (Le Poidevin 2004). Thus the block spacetime picture does not represent a realistic view of the real universe. The existence and uniqueness theorems underlying the usual block universe view (see Hawking and Ellis 1973), implying we can predict uniquely to both the future and past from any chosen time surface, do not apply to spacetimes including complex systems because the equations of state they assume are too simple - they don't include friction and dissipative effects, hierarchical structures, feedback effects, or the causal efficacy of information (Roederer 2005, Ellis 2006a), and they don't take quantum uncertainty into account. A better picture of the situation when realistic physics and biology is taken into account is provided by Evolving Block Universe proposal outlined above, in which the spacetime is seen as extending in a pointwise way as time evolves along all possible world lines.

The usual block view description is reasonably accurate for large scale space-times with very simple matter content (vacuum or barotropic equations of state for example), and so is acceptable for cosmological purposes or astrophysical studies such as collapse to a black hole. It is not adequate for small-scale descriptions of spacetime with complex matter or active agents such as living beings. It does represent the far future fate of the growing universe adequately, but in so doing does not imply that data on any spacelike surface enables one to predict or retrodict uniquely to later and earlier times. Rather it represents the quirky contingent nature of what actually happened in the real universe, which only became apparent as it unrolled.

## 5.1 Determinism and becoming

How do issues about determinism bear on Becoming? Does determinism entail a static block universe? Does indeterminism entail a dynamical or growing block universe? It seems that the first question has a negative answer, but the second one does not. The growing block universe could grow in either a deterministic or indeterministic fashion; the idea is compatible with both standard general relativity as expressed in the ADM theory, and with more realistic situations where the outcome of events is only determined as time unrolls (see section 2). Thus determinism does not necessarily imply a static block universe. On the other hand while a special relativity static block spacetime could contain indeterministic events, the events themselves in it would emerge as time progresses, so an EBU description would be appropriate for those events. But General Relativity implies the space time structure itself would then be changing. Through their gravitational effects, indeterministic events will change spacetime structure in a way determined by the outcome of those events, and imply a growing block universe picture is the appropriate one. In some circumstances these effects will be very small, nevertheless they will be there

How do issues of the time reversibility of the laws of physics bear on Becoming? A static block universe would seem to be compatible with both time reversibility and time irreversibility of laws, if spacetime is seen as the unchanging arena of physics; but as mentioned above, this is not the case when gravitational effects are taken into account. Then time irreversible laws imply a time-direction of 'becoming' at each event, which would seem to fit better with a growing block universe, because they go some way to demonstrating that the future and past are of a different character – which is what is made explicit in the idea of a growing block universe.

A common view is that Broad's idea of a growing block universe is unacceptable if `Becoming' is relativized to either a foliation of spacelike hypersurfaces or a family of timelike world lines. For example Gödel (1957) thought that such a relativized Becoming was not worth having: "The concept of existence … cannot be relativized without destroying its meaning completely." The viewpoint in this paper is that Becoming does indeed take place, and so physical theory had better recognise this feature. If the implication is a relativisation of time in relation to a foliation of timelike lines, so be it. This is something we will have to accept and live with.

### 5.2 The chronology protection conjecture

Does the picture presented here have any implications for the possible existence of closed timelike lines and associated causality violations? It is known that general solutions of the Einstein Field Equations do indeed allow such lines (see for example Hawking and Ellis 1973), but Hawking has proposed the ***chronology protection conjecture***: "*It seems that there is a Chronology Protection Agency which prevents the appearance of closed timelike curves and so makes the universe safe for historians*" (Hawking 1992, see also Visser 2002, Lockwood 2005, Hunter 2006).

An evolving block universe model, with potentiality transforming into existence as time progresses along each world line, may be a natural context for considering this question, providing a dynamic setting for considering constraints on the developing spacetime. One can prescribe an active form of the chronology protection conjecture in this setting, namely that as spacetime evolves along a set of physically chosen world lines, *it is forbidden that these world lines enter a spacetime domain that already exists*. This would provide the needed protection in a natural way.[25] There might be a cost in terms of existence of spacetime singularities in order that this can be accomplished in all cases (geodesic incompleteness might occur when this injunction is invoked); whether this is so needs investigation. But then the singularity theorems of general relativity theory (see Hawking and Ellis 1973) have often had this tension in them: in many cases, it is predicted that either a singularity occurs or there is a causality violation. There may additionally be a lack of uniqueness in the maximal causality-violation free extension; this needs investigation.

### 5.3 The arrow of time

If the EBU view is correct, the Wheeler-Feynman prescription for introducing the arrow of time by integration over the far future (Wheeler and Feynman 1945), and associated views comparing the far future with the distant past (Ellis and Sciama 1972, Penrose 1989), are not valid approaches to solving the arrow of time problem, for it is not possible to do integrations over future time domains if they do not yet exist. Indeed the use of half-advanced and half-retarded Feynman propagators in quantum field theory then becomes a calculational tool representing a local symmetry of the underlying physics that does not reflect the nature of emergent physical reality, in which that symmetry is broken.

---

[25] It has been said to me that quantum field theory assumes there is travel into the past, so causality conditions do not hold. My view on this is that Feynman diagrams with past directed world lines are a possible description of what happens, but there are preferable descriptions in which causality is compatible with special relativity in that all particle paths are future directed. In this case the time reversed diagrams express a symmetry of the system rather than the way the physics actually occurs.

The arrow of time problem needs to be revisited in this EBU context, with the collapse of the quantum wave function being a prime candidate for a location of a physical solution to the problem. We do not consider it further here.

**5.4 Issues of Ontology**
The hidden issue underlying all this discussion is the question of the ontological nature of spacetime: *does spacetime indeed exist as a real physical entity, or is it just a convenient way of describing relationships between physical objects, which in the end are all that really exist at a fundamental level*? Is it absolute or relational? Could it after all be an emergent property of interacting fields and forces (Laughlin 2005), or from deeper quantum or pre-quantum structure (Ashtekar 2005: Chapters 11-17)?

I will not pursue this contentious point here (for discussions, see e.g. Earman 1992, Hoeffer 1998, Huggett 2006). Rather I emphasize here that *the discussion in this paper is about models or representations of space time,* rather than making any ontological claims about the nature of spacetime itself. However I do believe that the kind of proposal made here could provide a useful starting point for a fresh look at the ontological issue, and from there a renewed discussion of the degree to which our representations of the nature of spacetime are an adequate representation of its true existential nature.

**Acknowledgement:**
I thank Bill Stoeger, two referees, and an editor for comments that have considerably improved this paper.

**References:**

[**Anninos 2001**] P Anninos, *Computational Cosmology: From the Early Universe to the Large Scale Structure*. http://relativity.livingreviews.org/Articles/lrr-2001-2/articlesu15.html , Section 6.1, "The Einstein Equations".

[**Arnowitt et al 1962**] R Arnowitt, S Deser and C W Misner, "The Dynamics of General Relativity". In *Gravitation: An introduction to Current Research*, Ed L Witten (Wiley, 1962), pp. 227-265.

[**Ashtekar 2005**] A Ashtekar, *100 Years of Relativity: Space time Structure: Einstein and Beyond*. (World Scientific, 2005).

[**Babcock and Collins 1929**] E B Babcock and J L Collins, "Does natural ionizing radiation control rate of mutation?". *Proc Nat Acad Sci* **15**: 623-628 (1929).

[**Bondi 1965**] H Bondi, "Some special solutions of the Einstein equations". In *Lectures in General Relativity*, *Brandeis Summer Institute in Theoretical Physics Vol I*, Ed. A Trautmann, F A E Pirani and H Bondi (Prentice Hall, 1965), 431-434.

[**Broad 1923**] C D Broad, *Scientific Thought* (New York: Harcourt, Brace and Co., 1923). For Table of Contents and some chapters, see http://www.ditext.com/broad/st/st-con.html .

[**Davies 1974**] P C W Davies, *The physics of time asymmetry*. (Surrey University Press, London, 1974).

[**Davies 2002**] P C W Davies, "That Mysterious Flow". *Scientific American* **287**, (September 2002), 40.

[**Dodelson 2003**] S Dodelson, *Modern Cosmology* (Academic Press, 2003).

[**Dunsby et al 1992**] P K S Dunsby, M Bruni and G F R Ellis, "Covariant perturbations in a multi-fluid medium". *Astrophys Journ* **395**, 54-74 (1992).

[**Earman 1992**] J Earman, *World Enough & Space-Time: Absolute vs. Relational Theories of Space & Time.* (Bradford Book, 1992).

[**Ellis 1971**] G F R Ellis, "Relativistic Cosmology". In *General Relativity and Cosmology*, Varenna Lectures, Course XLVII. Ed. R K Sachs (Academic Press, 1971), 104-179.

[**Ellis 1984**] G F R Ellis, "Relativistic cosmology: its nature, aims and problems". In *General Relativity and Gravitation*, Ed B Bertotti et al (Reidel, 1984), 215-288.

[**Ellis 2005a**] G F R Ellis, "Physics, Complexity, and Causality". *Nature* **435:** 743 (2005)**.**

[**Ellis 2005b**] G F R Ellis, "Physics and the Real World". *Physics Today* (July 2005), 49-54. Letters response, 59: 12-14 (March 2006).

[**Ellis 2006a**] G F R Ellis, "Physics and the real world". *Foundations of Physics,* April 2006, 1-36 [http://www.mth.uct.ac.za/~ellis/realworld.pdf].

[**Ellis 2006b**] G F R Ellis, "On the nature of emergent reality". In *The Re-emergence of Emergence*, ed P Clayton and P C W Davies (Oxford University Press, 2006) [http://www.mth.uct.ac.za/~ellis/emerge.doc].

[**Ellis and Buchert 2005**] G F R Ellis and T R Buchert, "The universe seen at different scales". *Physics Letter* A **347:** 38-46 (2005) [http://za.arxiv.org/abs/gr-qc/0506106].

[**Ellis and Sciama 1972**] G F R Ellis and D W Sciama, ``Global and non-global problems in cosmology". In *General Relativity*, ed. L. O'Raifeartaigh (Oxford University Press, 1972), 35-59.

[**Ellis and Williams 2000**] G F R Ellis and R M Williams, *Flat and Curved Space Times*. (Oxford University Press, 2000).

[**Feynman 1985**] R Feynman, *QED: The Strange Theory of Light and Matter.* (Princeton, 1985).

[**Fieser and Dowden 2006**] B Dowden, "Time". In *The Internet Encyclopaedia of Philosophy,* J Fieser and B Dowden (Eds), (2006). http://www.iep.utm.edu/t/time.htm .

**[Gödel 1957]** K Gödel, "A Remark About the Relationship between Relativity Theory and Idealistic Philosophy". In *Albert Einstein, Philosopher-Scientist, Vol. 2,* Ed. P A Schilpp, (Tudor Pub. Co, 1957).

[**Hadamard 1923**] J Hadamard, *Lectures on Cauchy's Problem in Linear Partial Differential Equations.* (New Haven: Yale University Press, 1923).

[**Hawking 1992**] S W Hawking, "The chronology protection conjecture". *Phys. Rev*. **D46**, 603-611 (1992).

[**Hawking and Ellis 1973**] S W Hawking and G F R Ellis, *The Large Scale structure of space time.* (Cambridge University Press, 1973).

[**Hinshaw 2006**] G Hinshaw, "WMAP data put cosmic inflation to the test". *Physics World* **19**, No 5 (May 2006), 16-19.

[**Hoefer 1998**]   C Hoefer**,** "Absolute versus Relational Spacetime: For Better or Worse, the Debate Goes On". *The British Journal for the Philosophy of Science* **49**(3): 451-467 (1998).

 [**Huggett 2006**] N Huggett, **"**The Regularity Account of Relational Spacetime". *Mind* **115** (457): 41-73 (2006).

[**Hunter 2006**] J Hunter, "Time Travel". In *The Internet Encyclopaedia of Philosophy* (2006), http://www.iep.utm.edu/t/timetrav.htm .

[**Isham 1997**] C J Isham, *Lectures on Quantum Theory, Mathematical and Structural Foundations*. (Imperial College Press, 1997).

[**Kolb and Turner 1990**] E W Kolb and M S Turner, *The Early Universe.* (Addison Wesley, 1990).

[**Laughlin 2005**] R B Laughlin, *A Different Universe: Reinventing Physics from the Bottom Down* (Perseus, 2005).

[**Le Poidevin 2004**] R Le Poidevin, "The Experience and Perception of Time", *The Stanford Encyclopedia of Philosophy (Winter 2004 Edition),* E N. Zalta (ed.), http://plato.stanford.edu/archives/win2004/entries/time-experience/ .

[**Lockwood 2005**] M Lockwood, *The Labyrinth of Time: Introducing the universe.* (Oxford University Press: 2005).

[**Markosian 2002**] N Markosian, "Time". The Stanford Encyclopedia of Philosophy (Winter 2002 Edition), E N. Zalta (ed.), http://plato.stanford.edu/archives/win2002/entries/time/ .

[**Mellor 1998**] D H Mellor, *Real Time II.* (Routledge, London: 1998).

[**Misner et al 1973**] C W Misner, K S Thorne and J A Wheeler, *Gravitation.* (Freeman, 1973).

[**National Academy 2005**] Committee on the origins and evolution of life, National Research Council: *The Astrophysical Context of Life* (Washington: National Academy Press: 2005) [http:// www.nap.edu/catalog/11316.html].

**[Penrose 1989]** R Penrose, *The Emperor's New Mind.* (Oxford University Press: 1989).

[**Percival 1991**] I Percival, "Schrödinger's quantum cat". *Nature* **351**, 357 (1991).

[**Roederer 2005**] J Roederer, *Information and its Role in Nature.* (Springer 2005).

[**Rothschild 1999**] L J Rothschild, "Microbes and Radiation". In *Enigmatic Micro-organisms and Life in Extreme Environments*, Ed. J Seckbach (Kluwer, 1999), p. 551.

[**Savitt 2001**] S Savitt, "Being and Becoming in Modern Physics", *The Stanford Encyclopedia of Philosophy (Spring 2002 Edition)*, Edward N. Zalta (ed.), http://plato.stanford.edu/archives/spr2002/entries/spacetime-bebecome/ .

[**Scalo et al 2001**] J Scalo, J Craig Wheeler and P Williams, "Intermittent jolts of galactic UV radiation: Mutagenetic effects" In *Frontiers of Life; 12th Rencontres de Blois*, ed. L. M. Celnikier (2001) [astro-ph/0104209].

[**Smart 1967**] J J C Smart, "Time". In *The Encyclopedia of Philosophy*, Ed. P Edwards (Collier-MacMillan, 1967), viii. 126-134.

[**Spiegel 1967**] M R Spiegel, *Theory and Problems of Theoretical Mechanics* (Schaum's Outline Series, McGraw Hill 1967), p. 65 and Ex. **3.19** on pp. 73-74.

[**Thomson and Stewart 1987**] J M T Thompson and H B Stewart, *Nonlinear Dynamics and Chaos*. (John Wiley and Sons, 1987).

[**Tooley 2000**] M Tooley, *Time, Tense, and Causation.* (Oxford University Press: Oxford, 2000).

[**Visser 2002**] M Visser, "The quantum physics of chronology protection". In *The Future of Theoretical Physics and Cosmology: Celebrating Stephen Hawking's 60th Birthday,* Ed. G. W. Gibbons, E. P. S. Shellard, S. J. Rankin.

[**Wald 1984**] R M Wald, *General Relativity.* (University of Chicago Press, 1984).

[**Wheeler and Feynman 1945**] J. A. Wheeler and R. P. Feynman, "Interaction with the Absorber as the Mechanism of Radiation". *Rev. Mod. Phys*. **17**, 157–181 (1945).

[**Zeh 1992**] H D Zeh, *The Physical Basis of the Direction of Time*. (Springer Verlag, Berlin, 1992).

---

version: 2006-08-01
(major revision: July 2006)